\documentclass[twocolumn]{scrartcl}
\usepackage{epsfig}
\usepackage{cite,latexsym,caption}
\usepackage[intlimits]{amsmath}
\newcommand{\abs}[1]{\lvert#1\rvert}
\begin{document}
\title{Isotope shift in the electron affinity of chlorine}
\author{Uldis Berzinsh, Morgan Gustafsson, Dag Hanstorp,\\
Andreas E.~Klinkm\"uller, Ulric Ljungblad\\
and Ann-Maire M{\aa}rtensson-Pendrill\\
Department of Physics, Chalmers University of Technology\\ 
and G\"oteborg University, \\SE 412\,96 G\"oteborg, Sweden}
\date{14 July 1994}
\maketitle
\begin{abstract}
  The specific mass shift in the electron affinity between $^{35}$Cl
  and $^{37}$Cl has been determined by tunable laser photodetachment
  spectroscopy to be $-0.51(14)$~GHz. The isotope shift was observed
  as a difference in the onset of the photodetachment process for the
  two isotopes. In addition, the electron affinity of Cl was found to
  be 29\,138.59(22)~cm$^{-1}$, giving a factor of 2 improvement in the
  accuracy over earlier measurements. Many-body calculations including
  lowest-order correlation effects demonstrates the sensitivity of the
  specific mass shift and show that the inclusion of higher-order
  correlation effects would be necessary for a quantitative
  description.
\end{abstract}
\begin{flushleft}
\end{flushleft}
%
\section{Introduction}
\label{int}
Tunable laser photodetachment spectroscopy makes high-resolution
studies of electron affinities possible, and in this work we
demonstrated an application of this technique to the measurement of
the isotope shift in the electron affinity of chlorine. The observed
shift is only a fraction of the \emph{normal mass shift} (NMS),
indicating a \emph{specific mass shift} (SMS) comparable to the normal
mass shift, but of opposite sign. The results obtained provide a
sensitive probe of electron corretation effects, which are known to be
particularly important for negative ions and the specific mass shift.

The outermost electron in a negative ion is very weakly bound: Far
from the nucleus it is bound only due to polarization of the other
electrons in the system, whereas the outermost electron in an atom or
positive ion is bound in a long-range Coulomb potential. Hence the
electron correlation is of greater importance in negative ions, which
therefore provide good tests of computational methods used to describe
electron correlation. In Li$^{-}$, e.~g., the electron correlation
energy has been found to be three times larger than the electron
affinity \cite{Sal-90}. Another system where the electron correlation
is dominating the binding energy is the two-electron system H$^{-}$,
which has been used as a model case in many calculations
\cite{Gel-90,Du-89,Vin-89}.

The specific mass shift arises from a correlation between the
electronic momenta through the motion of the nucleus with its finite
mass, and is discussed in more detail in section \ref{spe}. Being a
two particle operator the SMS is very sensitive to correlation
effects, and its calculation is a challenge for atomic theory. The
development of methods to include all-order correlation effects on the
SMS is in progress, but to our knowledge no calcualtion has yet been
performed for a many-electron system. Inclusion of the lowest-order
correlation terms in atomic many-body pertubation theory has been
found to give a significant---albeit insufficient---improvement in
several cases \cite{Mar-82,Hor-83,Har-91} and is here applied to the
electron affinity of chlorine, as described in section \ref{isoc}.

Since electron correlation is of major importance both for negative
ions and for the specific mass shift, it would be of particular
interest to investigate the specific mass shift in negative ions.
However, the only one experiment, so far, where an isotope effect has
been studied in an atomic negative ion is the measurement by Lykke et
al.~\cite{Lyk-91} of the electron affinities of hydrogen and
deuterium. For these light systems the mass shift is, of course, very
large, which enabled Lykke et al.~to observe a shift, although the
outgoing electron is a p wave, resulting in a slow onset of the
photodetachment process. The main reason for the lack of studies of
isotope shifts in negative ions is the absence of bound excited states
which are optically accessible. The only sharp structure in an optical
spectrum in the onset of the photodetchment process, and for most
negative ions, this threshold is in a wavelenght region not accessible
to tunable dye lasers. Schulz et al.~\cite{Sul-82} have applied
high-resolution techniques to study isotope effects in the negative
\emph{molecular} ions OH$^{-}$ and OD$^{-}$, where, however, the major
part of the mass effect is due to vibration and roation and not from
the electronic part, as in the case of atomic systems.

This work presents an investigation of the isotope shift of chlorine,
which was chosen for several reasons. From an experimental point of
view, Cl has the advantage, shared with all halogens, that an s
electron is emitted in the photodetachment, giving a sharp threshold.
In addition, it has two stable isotopes with large natural abundances.
From a theoretical point of view, the closed-shell ground state of a
halogen ion provides a convenient reference state for the caculations,
and makes the system relatively easy to treat. In addition, the
specific mass shift in the ionization potential of rare gases, wich
involve the same electron configuration, have been previously
investigated, theoretically as well as experimentally
\cite{Her-58,Ase-91,Wes-79}, and the comparison of results can give an
indication about the importance of various effects.
%
\section{Experimental procedure and results}
\label{exp}
\begin{figure*}
\begin{center}
\epsfig{file=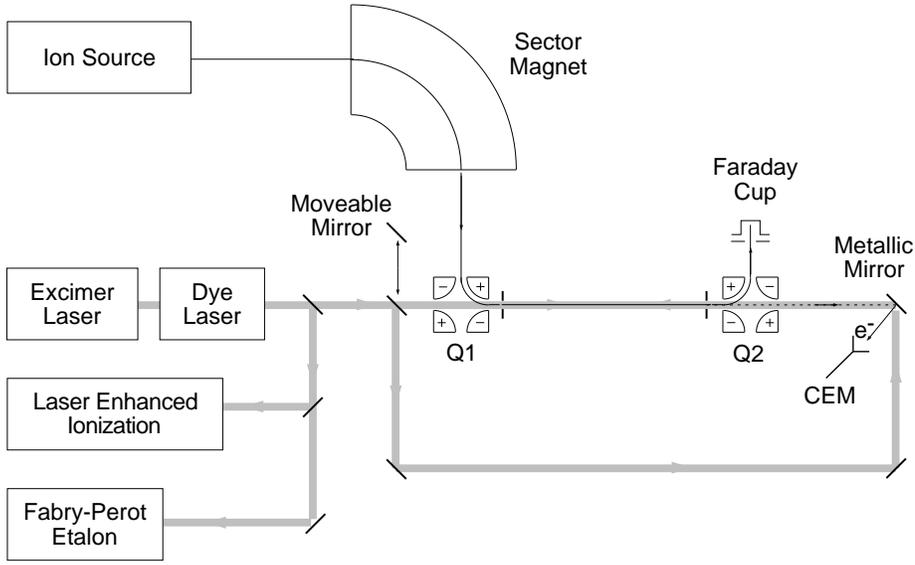,width=0.85\textwidth}
\caption{\label{kli-fig01}\sloppy Experimental setup.}
\end{center}
\end{figure*}

The experimental setup, shown in figure \ref{kli-fig01}, has been
described previously by Hanstorp and Gustafsson in connection with a
similar experiment where the electron affinity of iodine was
determnied \cite{Han-92-1}. In this experiment, negative chlorine ions
were formed from HCl gas on a hot LaB$_{\text{6}}$ surface. The ions
were accelerated to 3~keV, mass selected in a sector magnet and
focused by means of several Einzel lenses. In an analyzing chamber,
the ions were merged with a laser beam between two electrostatic
quadrupole deflectors placed 50~cm apart. An ion current of typically
6~nA for $^{35}$Cl and 2~nA for $^{37}$Cl was measured with a Faraday
cup placed after the second quadrupole deflector. The laser light
could be directed either parallel or antiparallel with respect to the
ion beam direction. Fast neutral atoms produced in the photodetachment
[rocess were not affected be the second quadrupole deflector. Instead,
they impinge on a glass plate coated with a conducting layer of tin
doped indium oxide (In$_{2}$O$_{3}$\,:\,Sn). Secondary electrons
produced by the neutral atoms were collected with a chanel electron
multiplier (CEM) operating in pulse counting mode. The CEM was
carefully shielded, both mechanically and electrically, to prevent
stray electrons from reaching the detector \cite{Han-92-2}.

%
An excimer pumped dye laser served as light source. The experiment was
done using two different laser dyes BMQ (Lambdachrome No.~3570) and PTP
(Lambdachrome No.~3400), with their peak intenities at 343~nm and 357~nm
respectively. The laser pulses, with a time duration of 20~ns, had an
energy of up to 1~mJ as measured in front of the experimental
chamber. Some of the laser light was directed through as Fabry Perot
etalon and through an acetylene-air flame for wavelength calibration.
%
\subsection{Detection of the photodetachment threshold}
\label{det}
According to the Wigner law \cite{Wig-48}, the cross section for
photodetachment of an s wave electron is, in the vicinity of the
threshold, given by the expression
\begin{equation*}
\sigma (E) = \begin{cases}
        c\sqrt{E-E_{0}},  &  E\ge
        E_{0}\quad , \\
        0, & E<E_{0}\quad 
        \end{cases}
\end{equation*}
where $\sigma (E)$ is the cross section for photodetachment, $E$ is
the photon energy, $E_{0}$ is the photodetachment threshold, and $c$
is a constant. The cross section behaviour is, however, complicated by
the hyperfine structure. The two chlorine isotopes both have a nuclear
spin of $\frac{3}{2}$. The ground state of the chlorine atom, which
has the configuration 3p$^{5}$\,$^{2}$P$_{3/2}$, is therefore split
into four hyperfine structure levels $F=0,1,2,3$, where the $F=0$
level has the lowest energy. The ground state configuration of the
negative chlorine ion, 3p$^{6}$\,$^{1}$S$_{0}$, shows neither fine nor
hyperfine structure. With one initial state and four possible final
states there are four independent photodetachment channels, which all
have to be included in the fitting procedure. The relative cross
section for the different channels is proportional to the multilicity
of the final state in the neutral atom, i.~e., proportional to $2F+1$,
and the total photodetachment threshold can therefore be described by
the function
\begin{multline*}
  \sigma (E) = c\sum_{F=0}^{3} (2F+1) \\
  \times \sqrt{E-\left( E_{0} + E^{F}_{\text{hfs}} \right)}\;,\quad
  E\ge E_{0}
\end{multline*}
where $E^{F}_{\text{hfs}}$ is the $F$ dependent energy correction due
to the hyperfine interaction \cite{Ful-76}, $E_{0}$ the threshold
energy for leaving the atom in the lowest hyperfine level of the
ground state of the atom, and $c$ is a constant.

\begin{figure}
\begin{center}
{\epsfig{file=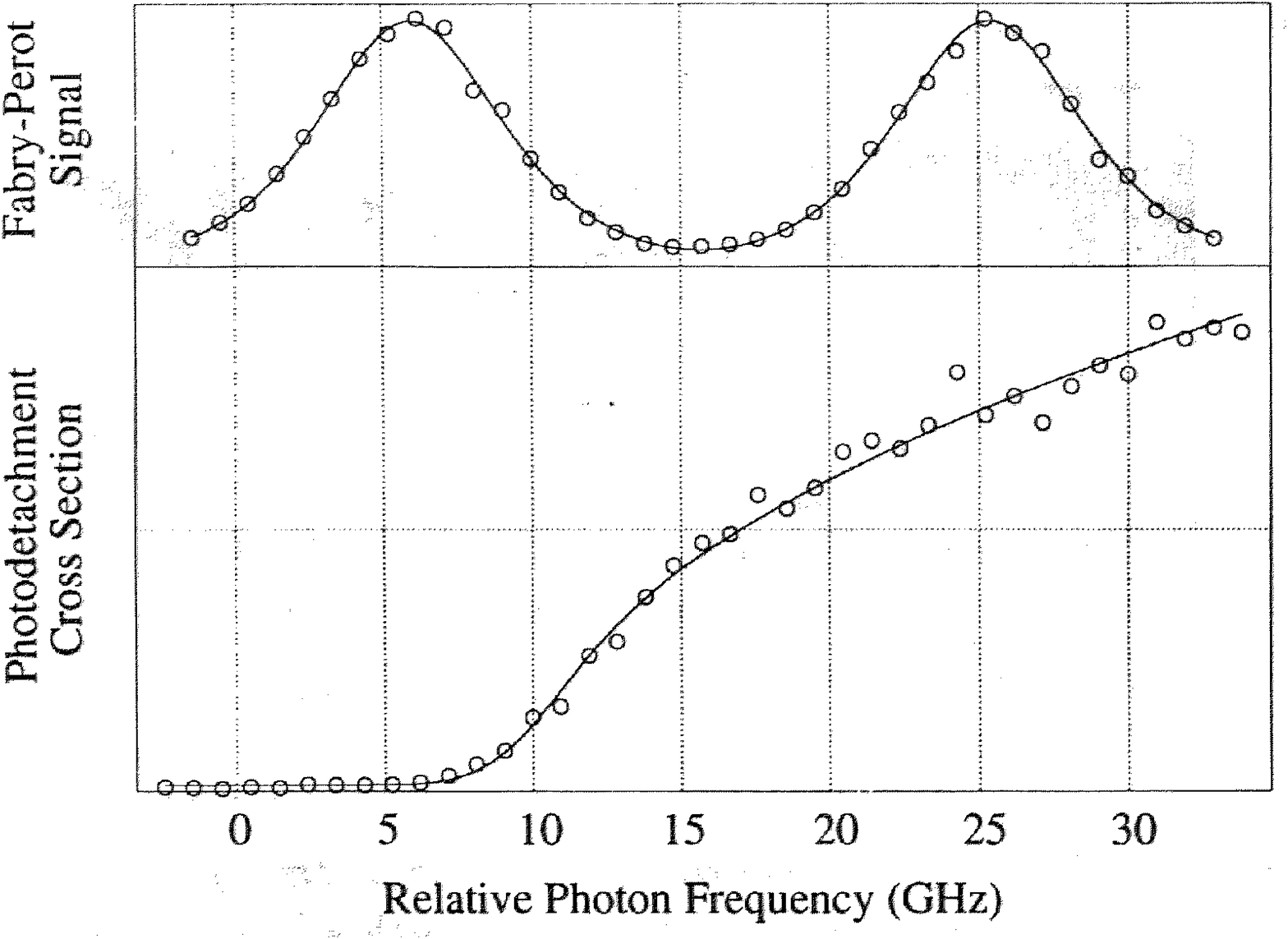, width=0.95\columnwidth}}
\end{center}
\protect\caption{\label{kli-fig02}\sloppy A scan used in the
  isotope-shift measurement. The lower curve is the relative cross
  section for photodetachment where the line is a fit to the
  experimental data. The upper curve is the transmission through a
  Fabry-Perot etalon used for frequency calibration.}
\end{figure}

The nonzero bandwidth $B$ of the laser also has to be taken into account
in the evaluation procedure. In our case, we can assume a Gaussian
intensity distribution $I(E^{\prime})$ with its peak at $E^{\prime}=0$
given by the expression \cite{Dem-88}
\begin{equation*}
I(E^{\prime}) = \frac{\text{e}^{-4(E^{\prime}/B)^{2}}}{2}\quad .
\end{equation*}
Including this bandwidth into the fitting procedure, the experimental
values can be fitted to the following modified Wigner law function:
\begin{multline}\label{eq04}
\sigma (E) = C \int_{E^{\prime}=-\infty}^{\infty} \sum_{F=0}^{3}
I(E^{\prime})(2F+1)\\
\times \biggl\{ \Bigl[ E + E^{\prime} - \bigl( E_{0} +
E_{\text{hfs}}^{F} \bigr)\Bigr]\\
+ \Bigl\lvert E+E^{\prime} -  \bigl( E_{0} +
E_{\text{hfs}}^{F} \bigr) \Bigr\rvert \biggr\}^{1/2}dE^{\prime} \quad .
\end{multline}
The term inside the absolute sign is included in order to make the
function equal to zero below the threshold. The experimental data
points were fitted to equation \eqref{eq04} using a nonlinear fitting
routine based on the Levenberg-Marquardt methode \cite{PFT-89}. An
example of such a fit is shown in figure \ref{kli-fig02}. Although
equation \eqref{eq04} is complicated, only the constant $C$ and the
threshold energy $E_{0}$ are varied to fit the data. The two Fabry
Perot peaks closest to the threshold, shown in the upper part of
figure \ref{kli-fig02}, were fitted to an Airy function by the same
numerical method, and their position served as frequency markers.

%
To determine the laser bandwidth, the expression in equation
\eqref{eq04} was fitted to one set of experimental data for different
values of the laser bandwidth. The standard deviation of those fits
vaired smoothly with energy, with a minimum at a bandwidth of
4.3~Ghz. This value agrees with the laser specification and was used for
all subsequent evaluations. 

In order to correct for the first-order Doppler shift, which is
sufficient in this experiment, the electron affinity
$E_{\text{EA}}(^{A}\text{CL})$ of the isotope $^{A}$CL is obtained as
average value of the energy of the photodetachment threshold,
$E_{\text{th}}$ obtained with parallel and antiparallel laser and ion
beams (designated p and a, respectively):
\begin{equation}\label{eq05}
E_{\text{EA}}(^{A}\text{CL}) = \frac{E_{\text{th}}^{\text{a}}
+ E_{\text{th}}^{\text{p}}}{2}\quad . 
\end{equation}
%
\subsection{Isotope shift}
\label{isos}
The isotope shift of the electron affinity in the negative chlorine
ion has been experimentally determined by measuring the difference in
electron affinity for the two stable chorine isotopes, each determined
as described in section \ref{det}.

\begin{figure*}
\begin{minipage}{\textwidth}
\parbox[b]{0.75\textwidth}{
\epsfig{file=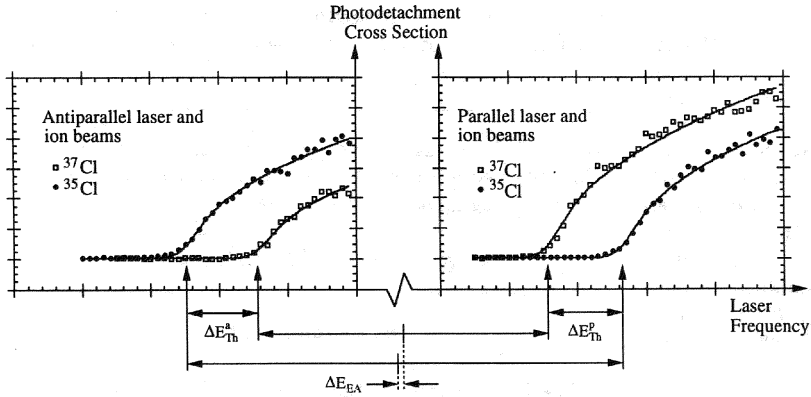, width=0.73\textwidth}
} \hfill
\parbox[b]{0.24\textwidth}{
\protect\caption{\label{kli-fig03}\sloppy\small A graphic description of
the evaluation of the isotope shift from four different experimentally
determind photodetahment thesholds.}  }
\end{minipage}
\end{figure*}

The frequency shift corresponding to the difference in electron
affinity, as shown in figure \ref{kli-fig03}, is given by the equation 
\begin{equation}\label{eq06}
\Delta E_{\text{EA}} = E_{\text{EA}} (^{37}\text{CL}) -  E_{\text{EA}}
(^{35}\text{CL})\quad .  
\end{equation}
Alternatively, this shift can be obtained by combining equations
\eqref{eq05} and \eqref{eq06} giving the expression
\begin{equation*}
\begin{split}
\Delta E_{\text{EA}} &= \frac{E_{\text{th}}^{\text{a}}
- E_{\text{th}}^{\text{p}}}{2} \\
&= \frac{E_{\text{th}}^{\text{a}}(^{37}\text{CL}) -
E_{\text{th}}^{\text{a}}(^{35}\text{CL}) }{2}\\
&{}\quad - \frac{E_{\text{th}}^{\text{p}}(^{35}\text{CL}) -
E_{\text{th}}^{\text{p}}(^{37}\text{CL}) }{2}\quad .
\end{split}
\end{equation*}
Experimentally, this has the advantage that two pairs of thresholds
situated very close to each other are compared, making the laser scans
much shorter. 

%
Four threshold measurements are thus required for one determination of
the difference in the electron affinities of the two isotopes, $\Delta
E_{\text{EA}}$. In the present work, we present a total of 13 isotope
shift deteminations, giving an experimental value
\begin{equation}\label{eq08}
\frac{\Delta E_{\text{EA}} }{h} = 0.34(14) \text{GHz}\quad .
\end{equation}
Since the experimtnal results are obtained by comparing the energy of
different thresholds, they are independent of an absolute wavelength
calibration. Many systematic effects will therefore cancel in the
evalutation procedure, giving a very high accuracy of the experimental
result. 

\begin{figure}
\begin{center}
{\epsfig{file=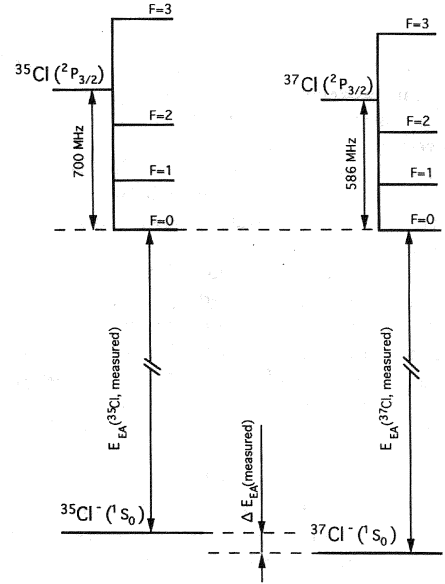, width=0.8\columnwidth}}
\end{center}
\protect\caption{\label{kli-fig04}\sloppy The energy level diagram of
the ground state of Cl$^{-}$ and Cl. The zero energy is set to the
lowest hyperfine level in the atomic ground state of the two isotopes.
}
\end{figure}

The electron affinity is defined from the threshold energy, and thus
involves the difference between the ground state of the negative ion and
the lowest hyperfine level of the gound state atom, whereas the isotope
shift is normally defined as the differnce between the fine-structure
levels unperturbed by the hyperfine interaction. In order to extract the
isotope shift from the experimental shift $\Delta E_{\text{EA}}$ in
equation \eqref{eq08}, it is necessary to correct for the energy
difference, $\Delta E_{\text{hfs}}^{F} = h\delta\nu_{\text{hfs}}^{F}$,
between the lowest hyperfine-structure level of the Cl atom and the
fine-structure level unperturbed by the hyperfine structure.  In
$^{35}$Cl, the $F=0$ ground-state level is situated
$\delta\nu^{F}_{\text{hfs}}(^{35}\text{Cl}) = 0.700$ GHz below the
unperturbed level and the corresponding shift in $^{37}$Cl is
$\delta\nu^{F}_{\text{hfs}}(^{37}\text{Cl}) = 0.586$ GHz. These shifts
are illustrated in figure \ref{kli-fig04} and were obtained using the
Casimir formula, where the $A$ and $B$ hyperfine constants were taken
from Fuller \cite{Ful-76}. Subtracting the difference between these two
values from $\Delta E_{\text{EA}}$ gives a corrected isotope shift of
the electron affinity of Cl:
\begin{equation}\label{eq09}
\begin{split}
\delta\nu_{\text{IS}} &= \frac{\Delta E_{\text{EA}}}{h} +
\delta\nu^{F}_{\text{hfs}}(^{37}\text{Cl}) -
\delta\nu^{F}_{\text{hfs}}(^{35}\text{Cl})\\
&=0.22(14)\;\text{GHz}\quad .
\end{split}
\end{equation}
%
%
\subsection{Electron affinity}
\label{ele}
In addition to the isotope-shift measurements, an improved absolute
value of the electron affinity of $^{35}$Cl has been determined.
Extended descritions of the experimental procedure for this experiment
has been given elsewhere \cite{Han-92-1,Han-93}.

\begin{figure}
\begin{center}
{\epsfig{file=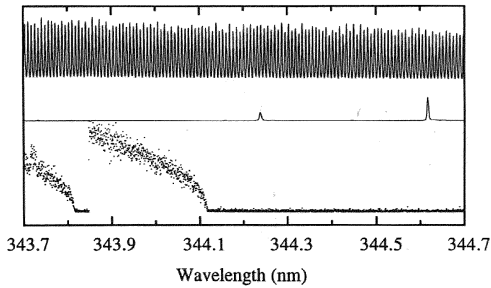, width=0.9\columnwidth}}
\end{center}
\protect\caption{\label{kli-fig05}\sloppy A complete recording of a
  spectrum used for determination of the electron affinity. The lower
  curve is the relative cross section, the middle curve is
  laser-enhanced ionization of nickel in a flame, and the upper curve
  is the transmission through a Fabry-Perot {\'e}talon. The direction
  of the laser beam was reversed at approximately 343.85~nm.  }
\end{figure}

The laser wavelength was scanned between 343.7$\dots $344.7~nm in
steps of 0.000\,5~nm. At each wavelength the signal from 30 laser
pulses were averaged. In total seven scans, one of which is shown in
figure \ref{kli-fig05}, were recorded. Three different signals were
registered, namely, the number of atoms neutralied in the
photodetachment process, laser enhanced ionization of nickel in an
acetylene-air flame \cite{Axn-84}, and the transmission through a
Fabry Perot etalon. The direction of the laser beam was reversed in
the middle of the scans in order to aquire the threshold position for
both the parallel and the antiparallel laser-ion beam geometries in
each scan. The threshold positions were evaluated using the method
described in section \ref{det}. The resonant absorption lines in
nickel were fitted to Lorentzian functions in order to find their peak
positions. The Fabry Perot etalon fringes served as frequency markers
in order to connect the threshold energies with the reference lines.
The electron affinity, corrected for the Doppler shift, was determined
by taking the mean value of the energies for the parallel and
antiparallel photodetachment thresholds, as shown in equation
\eqref{eq05}.

The energies of the nickel reference lines have recently been measured
using Fourier-transform spectroscopy \cite{Lit-93}, and they are known
to within $10^{-3}$~cm$^{-1}$. As previously mentioned, we use
laser-enhanced ionization in a flame in order to detect these lines
\cite{Axn-84}. The main sources of uncertainty in our calibration
procedure are broadening and shifts caused by collisions in the flame.
Spectral lines are at the most redshifted with 1/3 of the pressure
broadening, and experimentally observed shifts are normally smaller
\cite{Wag-73}. The widths of the spectral line profiles that we
observed were approximately 11~GHz. When the contribution from the
laser bandwidth and the Doppler broadening has been taken into
account, a remaining pressure broadening of 10~GHz was obtained. We
therefore estimate the redshift of our reference lines to be smaller
than 3.3~GHz. Isotope shifts of the nickel reference lines have been
measured \cite{Sro-61} but are too small to influence our
measurements.

The statistical scattering in this measurement is only
0.013~cm$^{-1}$. This corresponds to 0.4~GHz, which is less than one
tenth of the laser bandwidth. In the error in our final result of the
electron affinity, we include two standard deviations of the
statistical uncertainty plus the maximum possible systematic shift due
to the calibration procedure. This results in an electron affinity of
\begin{equation*}
\frac{E_{\text{EA}}}{hc}= 29\,138.59(22)\;\text{cm}^{-1}\quad , 
\end{equation*}
which is within the uncertainty of the previously reported most
accurate value 29\,138.3(5)~cm$^{-1}$ \cite{Tra-87}, and an
improvement of the accuracy with a factor of two. Further, it is
within the uncertainty of the result of 29\,138.9(7)~cm$^{-1}$ which
we recently reported \cite{Han-93}. In the present investigation, we
have used more accurately known atomic reference lines which lie
closer to the photodetachment thresholds, and we have improved the
statistics. The Ni reference lines used in the investigations are
(3d$^{8}$4s$^{2}$\,$^{3}$F$_{4}$ $\mapsto$
3d$^{8}$4s4p\,$^{5}$F$_{4}$) at 29\,084.455(1)~cm$^{-1}$ and
(3d$^{9}$4s\,$^{3}$D$_{3}$ $\mapsto$ 3d$^{9}$4p\,$^{3}$F$_{3}$) at
29\,115.975(1)~cm$^{-1}$. All these values of the electron affinity of
chlorine agree within their uncertainties, but they are outside the
error bars of the value of 29\,173(24)~cm$^{-1}$ recommended by Hotop
and Lineberger in their compilation of atomic electron affinities
\cite{Hot-85}. With three additional determinations of the electron
affinity of chlorine, it should be possible to give a more accurate
recommended value.
%
\section{Isotope shift calculations}
\label{isoc}
The different nuclear masses and charge distribution of $^{35}$Cl and
$^{37}$Cl leads to a chift in the electron affinity which can be
written as \cite{Kin-84}
\begin{equation}\label{eq10}
\begin{split}
\delta\nu^{A,A^{\prime}}_{i} = K^{\text{MS}}_{i} \frac{M_{A} -
M_{A^{\prime}}}{(M_{A^{\prime}} + m_{\text{e}})M_{A}}\\
{}+F_{i}(1-\kappa)\delta\langle r^{2} \rangle ^{A,A^{\prime}}\; ,
\end{split}
\end{equation}
where the first term is the mass shift accounting for the nuclear motion
and the second term is the field shift arising from the change in
electrostatic potential from the nucleus due to the change in charge
distribution. For light elements, like Cl, the mass shift dominates. It
falls off rapidly with the nuclear mass, whereas the field shift
increases for heavy elements, since the orbitals become more and more
contracted and the nucleus more extended. The small correction for the
field isotope shift is discussed in section \ref{fie}.

The expression for the mass shift factor is found by expanding the
nuclear kinetic energy in equation \eqref{eq10} as
\begin{align*}
\frac{\boldsymbol{P}^{2}_{\text{N}}}{2M_{A}} &=
\frac{(\sum_{i}\boldsymbol{p}_{i})^{2}}{2M_{A}}\\
&=\frac{\sum_{i}\boldsymbol{p}^{2}_{i} + \sum_{i\neq
j}\boldsymbol{p}_{i}\cdot\boldsymbol{p}_{j}}{2M_{A}}\quad .
\end{align*}
We find that the mass shift factor can be divided into two terms,
\begin{align*}
K^{\text{MS}} &= K^{\text{NMS}} + K^{\text{SMS}}\\
\intertext{where the first term}
K^{\text{NMS}} &= m_{\text{e}} \nu
\end{align*}
gives the ``normal'' mass shift, and accounts for the effect of
substituting the electron mass
\begin{align*}
m_{\text{e}} &= 5.485\,8\times 10^{-4}\;\text{u}\\
\intertext{by the reduced mass} 
\mu &= \frac{m_{\text{e}} M}{m_{\text{e}} + M}
\end{align*}
in the Schr{\"o}dinger equation. For the electron affinity of Cl,
\begin{equation*}
K^{\text{NMS}} = 479.43\;\text{GHz u}\quad .
\end{equation*}
The mass shift between the two Cl isotopes is obtained by multiplication
of the mass shift factor, $K^{\text{MS}}$, with the mass factor
\begin{equation*}
\frac{M_{37} - M_{35}}{M_{35}(M_{37} + m_{\text{e}})} \approx
0.001\,55\;\text{u}^{-1}\quad ,
\end{equation*}
giving
\begin{equation*}
\delta\nu^{\text{NMS}} = 0.74\;\text{GHz}\quad .
\end{equation*}
The nuclear masses
\begin{align*}
M_{35} &= 34.96\;\text{u}\\
\intertext{and}
M_{37} &= 36.96\;\text{u}
\end{align*}
are obtained by subtracting 17 $m_{\text{e}}$ from the atomic masses.

The second part,
\begin{equation*}
K^{\text{SMS}} = \frac{\Delta\langle\sum_{i\neq j}
\boldsymbol{p}_{i}\cdot\boldsymbol{p}_{j}\rangle}{h}\quad ,
\end{equation*}
involves two electrons simultaneously and describes a correlation
between the electronic momenta arising through the motion of the
nucleus. Here, we use this nonrelativitic operator, although the
orbitals used for the evaluation were obtained relativistically. The
theoretical results are presented in terms of $K^{\text{SMS}}$ in
order to facilitate comparison between results for different elements.
%
\subsection{The specific mass shift}
\label{spe}
The observable shift in the electron affinity results from the change
in this expectation value between the closed shell ground state of
Cl$^{-}$ and the ground state of the neutral system with its
3p$_{3/2}$ hole. This can be described using essentially the same
formalism as developed for systems with one valence electron outside a
closed shell, although the correlation between a hole and the core
tends to be more important. In this work, however, we have included
only lowest order correlation effects, using direct summation of
numerical relativistic basis functions obtained using the methods
described by Salomonson and {\"O}ster \cite{Sal-89}. This method has
been applied to SMS calculations for Cs, Tl \cite{Har-91}, and
Yb$^{+}$ \cite{Mar-94} and the generalization to a hole follows that
used in the nonrelativistic calculation for the shift of the binding
energy of Ne \cite{Hor-83}.

For scalar operators, like the SMS, the core gives very large
contributions ($\approx -144~\text{THz u}$ in this case), which,
however, cancel between initial and final states. An advantage of
pertubation theory is the cancellation of contributions from the
unperturbed core automatically obtained by omitting the ``zero-body''
terms describing these contributions. In the case of a single hole
state only ``one-body'' terms need to be evaluated.

The first contribution to $K^{\text{SMS}}$ is the interaction of the
hole with all core electrons. Only the exchange terms contribute to the
expexctation value, making the first-order value negative. Evaluating
the expectation value between Dirac-Fock (DF) orbitals for the ground
state of Cl$^{-}$, which is used as a refernece, gives $-3.68\;\text{THz
u}$. The DF value is a sum over the interactions of the 3p$_{3/2}$ with
all core orbitals whose $l$-quantum number differs by one unit and $j$
value differs by at most one unit from the value of the hole, i.e., all
the $n$s electrons. The ineraction with 1s electrons is found to
dominate, giving $-2.24\;\text{GHz u}$---the smaller overlap is
compensated by the larger momenta of the inner core electrons.

The ionization of Ne, studied in earlier work \cite{Hor-83}, involves
the removal of a 2p$_{3/2}$ electron. In that case, the DF contribution
to the mass shift constant, $K^{\text{SMS}}$, is $-9.00\;\text{THz u}$,
of which the 1s-2p interaction accounts for $-8.24\;\text{THz u}$. In
spite of the larger nuclear charge of Cl, which leads to increased
electronic momenta, the higher principal number of the active electron
in Cl leads to a smaller DF value for $K^{\text{SMS}}$.

\begin{figure}
\begin{center}
{\epsfig{file=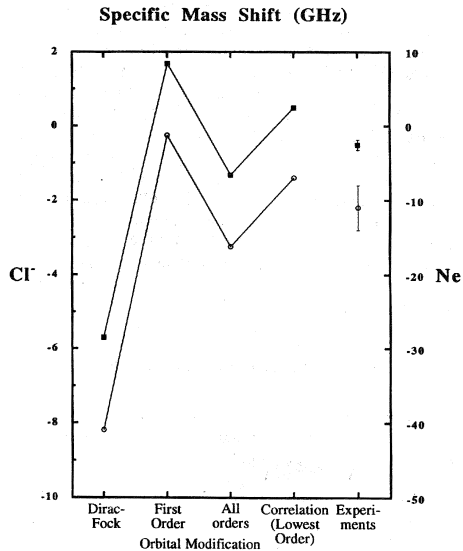, width=0.85\columnwidth}}
\end{center}
\protect\caption{\label{kli-fig06}\sloppy\small A comparison
  between the experimental value for the specific mass shift for the
  electron affinity in Cl and theoretical results at different levels
  of approximation. For comparison, the corresponding results are also
  shown for the shifth of the ionization potential of Ne (with the
  scale to the right). The first value is evaluated using orbitals
  obtained in the Dirac-Fock potential from the unperturbed
  closed-shell core and the next two values show, respectively, the
  lowest-order and all-order effect of the modification of the core
  orbitals due to the removal of an $n$p electron. The final
  theoretical value accounts also to lowest (i.e., second) order for
  the corretation effects and involves the simultaneous excitation of
  two electrons (GHz).  }
\end{figure}

The nuclear motion affects the wave function of \emph{all} electrons,
thereby modifying, e.g., the interaction between the core and the 3p
hole. The change in the 3p energy due to the first-order correction of
the core gives a contribution $4.77\;\text{THz u}$ to the SMS constant,
i.e., larger than the DF value but with oppsite sign. However, these
orbital corrections affect also the interactions within the core. A
self-consistent treatment of the core orbital modifications, closely
related to the ``random-phase approximation'' (RPA) approach, reduces
the orbital modification effect by about 40 \%, to $2.83\;\text{THz
u}$. Adding this value to the DF contribution cancels a large part of
it, leaving a sum of only $-0.84\;\text{Thz u}$. We note that Bauche
\cite{Bau-74}, in his pioneering nonrelativistic isotope-shift
calculations, performed seperate Hartree-Fock calculations for the
initial and final states and in this way include automatically the
``RPA'' terms. Also for the shift of the ionization energy in Ne, the
orbital modifications give significant reduction of the mass shift
constant, athough the corrections for an atom are not quite as drastic
as for a negative ion, where the interaction between the electrons plays
a more dominant role. The first-order modification gives
$8.72\;\text{THz u}$, changed to $5.43\;\text{THz u}$ by the
self-consistent treatment. Figure \ref{kli-fig06} shows the theoretical
results at different levels of approximation for these systems, together
with the experimental values.
%
Being a two-paricle operator the SMS is very sensitive to correlation
effects, which arise already in second order where they were found to
give $1.7\;\text{THz u}$, thereby changing the overall sign, giving
\begin{equation*}
K^{\text{SMS}} = 0.32\;\text{THz u} 
\end{equation*}
corresponding to a shift between the two stable Cl isotopes of
$0.50\;\text{GHz}$. In view of the large cancelations, this result is
very uncertain. For Ne, the lowest-order correlation effect,
$2.04\;\text{THz u}$, leads to a theoretical value $-1.54\;\text{THz
  u}$, compared to the experimental value $-2.4\;\text{THz u}$.
Higher-order effects must thus reduce the correlation contribution for
Ne. Lowest-order correlation effects, in fact, often give an
overestimate, as pointed out by Dzuba et al.~\cite{Dzu-88}, who
advocate the use of a screened Coulomb potential to describe the
electron correlation. The modification of the valence orbital to an
approximate Brueckner orbital is often a very improtant correlation
effect \cite{Lin-76-3}. Here, we evaluated these corrections to
the lowest order, and found a reduction of the DF and RPA results of
only about 1.4 \%. A more complete treatment of higher-order
correlation effects would be essential to obtain agreement with
experimental data.
%
\subsection{The field isotope shift}
\label{fie}
In addition to the mass-dependent isotope shift decribed above, a
shift in the electron affinity can also arise due to changes
$\delta\langle r^{2}\rangle ^{35:37}$ in the nuclear charge
distribution, as shown in equation \eqref{eq10}. The electronic factor
$F$ for the field shift is given by
\begin{equation*}\label{eq11}
F=-4\pi\Delta \abs{\Psi(0)}^{2} \frac{Z}{6}\frac{q_{\text{e}}^{2}}{4\pi
\epsilon_{0}}
\end{equation*}
and is determined by the change in electron density, $\abs{\Psi
  (0)}^{2}$, at the nucleus between the lower and upper state of the
transition. The 3p$_{3/2}$ electron, itself, has a negligible density
at the nucleus. Nevertheless, its removal causes a change in the
distribution of the s electrons in the core. Including the RPA-type
corrections to first order gives
\begin{equation*}
F\approx 254\;\text{MHz/fm}^{2}\quad .
\end{equation*}
Similar to the SMS, a self-consitent treatment of the RPA terms gives
a significant reduction:
\begin{equation*}
F\approx 117\;\text{MHz/fm}^{2}\quad . 
\end{equation*}
We note
that this value is comparable to the factor
\begin{equation*}
F_{\text{4s}}(\text{K})\approx -103\;\text{MHz/fm}^{2} 
\end{equation*}
for the binding energy of the 4s ground state of the nearby
alkali-metal K, but is considerable larger than the factor for its
first p state
\begin{equation*}
F_{\text{4p}}\approx 4.6\;\text{MHz/fm}^{2}\quad .
\end{equation*}

The correction term $\kappa $ in equation \eqref{eq10} accounts for
higher moments, $\delta\langle r^{4}\rangle$, $\delta\langle
r^{6}\rangle$, $\ldots$, of the nuclear charge distribution, but is
negligibly small (about 0.4\%) for Cl.

Briscoe et al.~\cite{Bri-80} have measured the charge distribution of
$^{35}$Cl and $^{37}$Cl. using electron scattering. From their
parameters for the Fermi distribution, we obtain
\begin{equation*}
\delta\langle r^{2}\rangle ^{35:37} = 0.12(12)\;\text{fm}^{2}\quad .
\end{equation*}
Combining this result with or calculated $F$ value indicates that the
field isotope shift of the electron affinity
\begin{equation*}
\delta\nu_{\text{FS}}^{35:37} = 14(14)\;\text{MHz}\quad ,
\end{equation*}
which is well below the error bars of the isotope shift measurement.
%
\section{Comparison between theory and experiment}
\label{com}
The experimentally determined difference between electron affinity for
the naturally occuring chlorine isotopes of the fine structure levels
unperturbed by the hyperfine interaction was $\Delta\nu_{\text{IS}} =
0.22(14)\;\text{MHz}$, as disccued in section \ref{isos}.

Subtraction of the normal mass shift (see section \ref{isoc})
\begin{equation*}
\delta\nu_{\text{NMS}} = \frac{\nu m_{\text{e}}\Delta M}{M_{35}M_{37}} =
0.74\;\text{GHz}
\end{equation*}
give a ``residual shift'' of $-0.51(14)\;\text{GHz}$. As seen in section
\ref{fie}, the field shift is only
\begin{equation*}
\delta\nu_{\text{FS}} = 14(14)\;\text{MHz}\quad ,
\end{equation*}
leaving a specific mass shift of
\begin{equation*}
\delta\nu_{\text{SMS}} = -0.51(14)\;\text{GHz}
\end{equation*}
corresponding to a factor
\begin{equation*}
K_{\text{SMS}} = -0.33(9)\;\text{THz u}\quad .
\end{equation*}
This value is close to the theoretical result, but with opposite sign.

In order to understand better the cause of this apparent
contradiction, we show in figure \ref{kli-fig06} the results obtained
at various levels of approximation. The final result arises from large
cancellation. The DF expectation value is many times larger than the
final result. The lowest-order orbital modification (``RPA terms'') is
of comparable size and with opposite sign. Higher-order RPA terms
cancel about 40\% of the lowest-order orbital modification crrection.
It seems like the lowest-order correlation effects are overestimated
by a similar amount: the experimental value is about halfway between
the result with and without lowest-order correlation. For comparison
we shown in figure \ref{kli-fig06} also the experimental \cite{Wes-79}
and analogous theoretical values \cite{Hor-83} for the specific mass
shift in the ionization potenital of neon. The calculated specific
mass shift for neon is $-7\;\text{GHz}$, whereas the experimental
value is $-11.3\;\text{GHz}$. Although slighlty outside the
uncertainty of the experimental value, the theoretical value in this
case has the right sign. The negative first-order DF value plays a
more dominant role for an atom than for a negative ion.

Similar results are obtained for the chlorine electron affinity,
itself: The DF value, 0.1480~a.u., is about 11\% larger than the
experimental value 0.1328~a.u., but the lowest-order correlation
effects bring a reduction of about 20\%, giving a theorectical value
of 0.1181~a.u., again demonstrating the need of more complete
treatment of correlation effects. The coupled-cluster approach
provides a systematic procedure for including higher-order effects
\cite{Lin-78}, which has given quite accurate results for systems with
one valence electron \cite{Sal-91}, and could be applied also to hole
states, although that has not yet been implemented in our program.
%
\section{Conclusion}
\label{con}
We have determined the isotope shift in the electron affinity of Cl to
be 0.22(14)~GHz, of which $-0.51(14)$~GHz is due to the specific mass
shift. This is to our knowledge the highest resolution obtained in any
laser experiment on atomic negative ions, and the resolution is even
higher than in electron affinity determinations where single-mode dye
lasers have been used \cite{Neu-85}. This could be achieved since an
isotope-shift measurement involves comparison of photodetachment
thresholds, rather than absolute determination. Many systematic
uncertainties will therefore cancel, giving a very high accuracy in
the experimental result. Further, the collinear geometry with the high
photon flux from the pulsed dye lasers gives very good counting
statistics.

Experimentally it would be very hard to improve the resolution on the
isotope shift on chlorine since the measured shift is only $1/30$ of
the bandwidth of the laser. A more precise experiment could, however,
be made using a negative ion where the photodetachment threshold
corresponds to the wavelength region of a single-mode dye laser. The
laser light should then be amplified in a pulsed amplifier in order to
keep the good signal to background ratio obtained in the present
experiment. $\text{S}^{-}$ could then be a suitable choice, although
it would probably be necessary to use isotopically enriched sulfur.
The theoretical treatment of such a system is more complicated, but
might be amenable to multiconfiguration Hartree-Fock calculation
\cite{Fis-91,Sue-92}. Investigating a few-electron system, like
$\text{Li}^{-}$, would give more insight, since very accurate
calculations could be performed. In this case, however, the
experimental conditions are much more unfavorable. Nevertheless, the
experimental accuracy in the shift of the electron affinity of Cl is
sufficient to demonstrate the importance of higher-order correlation
effects.

Finally, we conclude that this is the third experiment showing an
electron affinity of chlorine deviating from the currently recommended
value. A more precise recommendation of this quantity could therefore
be given by means of this and previous experiments.
%
\section*{Acknowledgements}
\label{ack}
We would like to express our appreciation to Ove Axner, who lent us
the laser system used in the experiment. Financial support by the
Swedish Natural Science Research Council is gratefully acknowldged.
The visit by Uldis Berzinsh was made possible by financial support
from the Swedish Institute and from the International Research Fund at
G{\"o}teborg University.
\clearpage
\onecolumn
\bibliography{physjabb,qualli}
\end{document}